\documentclass[aps,pra,pifont,citesort,twocolumn]{revtex4}
\usepackage{lineno}
\usepackage{color,soul}
\usepackage{setspace}
\usepackage[utf8]{inputenc}
\usepackage{graphicx,amsmath,amssymb}
\usepackage{dcolumn}
\usepackage{bm}
\usepackage{times}
\usepackage{mathtools}

\newcommand{\bs}[1]{{\boldsymbol #1}}
\newcommand{\erf}[1]{Eq.~(\ref{#1})}
\newcommand{\erfs}[2]{Eqs.~(\ref{#1}) and (\ref{#2})}

\newcommand{\ket}[1]{\left |#1\right \rangle}

\newcommand{\nn}{\nonumber}

\definecolor{darkgreen}{rgb}{0.0, 0.42, 0.24}
\definecolor{orange}{rgb}{1.0, 0.55, 0.00}

\newcommand{\blk}{\color{black}}

\newcommand{\grn}{\color{darkgreen}}

\usepackage{epstopdf} 
\usepackage{epsfig}
\begin{document}

\title{Reference frame independent Einstein-Podolsky-Rosen steering }
\author{Sabine Wollmann}
\author{Michael J. W. Hall}
\author{Raj B. Patel}
\author{Howard M. Wiseman}
\author{Geoff J. Pryde}

\email{g.pryde@griffith.edu.au}

\affiliation{
Centre for Quantum Dynamics, Griffith University, Brisbane, Queensland 4111, Australia}

\date{\today}

\begin{abstract}

Protocols for testing or exploiting quantum correlations---such as entanglement, Bell nonlocality, and Einstein-Podolsky-Rosen steering--- generally assume a common reference frame between two parties. Establishing such a frame is resource-intensive, and can be technically demanding for distant parties. While Bell nonlocality can be demonstrated with high probability for a large class of two-qubit entangled states when the parties have one or no shared reference direction, the degree of observed nonlocality is measurement-orientation dependent and can be arbitrarily small. In contrast, we theoretically prove that steering can be demonstrated with $100\%$ probability, for a larger class of states, in a rotationally-invariant manner, and experimentally demonstrate rotationally-invariant steering in a variety of cases. We also show, by comparing with the steering inequality of Cavalcanti \textit{et al.} [{J. Opt. Soc. Am. B} \textbf{32}, A74 (2015)],  
that the steering inequality we derive is the optimal rotationally invariant one for the case of two settings per side and two-qubit states having maximally mixed reduced (local) states.

\end{abstract}
\maketitle

Shared quantum correlations are a topic of significant foundational interest, and an important resource for 
quantum information and communication protocols. Quantum steering (also known as Einstein-Podolsky-Rosen (EPR) steering) corresponds to a class of correlations stronger than those required to merely witness entanglement, but which need not violate any Bell inequality ~\cite{Wiseman2007}. 
Moving down this hierarchy of correlation strength, from Bell nonlocality to steering to entanglement, gives access to protocols which are more robust to noise~\cite{Saunders2010, Bennet2012}. The cost is that, while Bell inequality violations require neither party (Alice or Bob) to be trusted, steering requires one (here, Bob) to be trusted, and regular entanglement witnessing requires full trust in both parties \cite{referred}. Steering therefore represents an interesting and important case, providing for strong~\cite{Bennet2012,Smith2012}, even loophole-free~\cite{Wittmann2012}, tests of nonlocality, but without the extreme noise suppression required to achieve Bell inequality violations. 

In general, all these correlation tests, and the quantum information tasks that derive from them, assume a shared reference frame between the parties, Alice and Bob. Establishing such a common reference frame is a nontrivial issue in experimental situations. For instance, in quantum communication, a time varying temperature can change the orientation of the polarisation reference frame in optical fibre. Likewise, the relative measurement settings between a satellite and earth could be time-varying. In both cases, active compensation of these changes presents a considerable challenge ~\cite{Laing2010}.
Such compensation becomes unnecessary if encoding in optical orbital angular momentum ~\cite{DAmbrosio2012} or complicated entangled states~\cite{Cabello2003}. However, such states are very susceptible to loss and noise, and generating and manipulating such systems may be difficult. Therefore, it is of interest to reduce reference-frame dependence
in quantum information tasks.

Can nonlocality be demonstrated simply without having established a common reference frame? This question was recently answered theoretically~\cite{Liang2010,Laing2010,Wallman2011} and experimentally~\cite{Shadboldt2012,Palsson2012} for Bell nonlocality.
Here, we demonstrate that a quantum steering protocol between two parties can be performed without establishing a reference frame. We can contrast our results with the case for Bell violations, which are measurement-orientation dependent and can be arbitrarily small---our technique surpasses these limitations.

To investigate quantum steering without a reference frame, we derived and experimentally tested a new, rotationally-invariant steering (RIS) inequality, which is very robust and can certify steering with 100\% probability for a large class of two-qubit entangled states. We compare the case where the parties share one measurement direction (e.g., derived from line of sight between them or the propagation axis of an optical fibre) to the case where they share none. 
We demonstrate the further advantage obtained by increasing the number of measurement settings each party uses.
 We compare our new RIS inequality with the steering inequality of Cavalcanti \textit{et al.}~\cite{Cavalcanti2015}. This allows us to test and prove the optimality of our RIS inequality when there are two measurement settings per side and their marginal distributions are maximally mixed.

{\it Rotationally invariant steering (RIS) inequalities.--} 
A general quantum steering protocol between two parties, Alice and Bob, proceeds as follows. In each round, Bob receives a quantum system and announces two randomly chosen measurement settings: $j\in \{1,\dots,m\}$ for Alice; and  $k\in\{1,\dots,n\}$ for himself (in many previous realisations $j=k$). 
 Alice announces a corresponding measurement outcome $A_j$, which may be the result of a genuine measurement on her half of an entangled pair that she shares with Bob, or the result of a strategy that she (or some adversary controlling her equipment) is using to try to cheat, i.e. to convince Bob of shared quantum correlations which do not exist. Bob measures a pre-agreed observable $\hat B_k$ on Hilbert space $H_B$, with outcome $B_k$. Over many runs Bob is able to estimate the correlation matrix $M_{jk}:=\langle A_jB_k\rangle$, and test whether it is compatible with a local hidden state (LHS) model for his system, i.e., a set of states $\{\hat\varrho_\lambda\}$ on $H_B$ such that 
\begin{equation} \label{lhs}
M_{jk}= \langle A_jB_k\rangle =\int d\lambda \,p(\lambda) \,\langle A_j\rangle_\lambda\,\langle \hat B_k\rangle_{\hat\varrho_\lambda}.
\end{equation}
Here $\lambda$ labels an underlying variable with probability 
density $p(\lambda)$, $\langle \hat B_k\rangle_{\hat\varrho_\lambda}:={\rm Tr}[\hat\varrho_\lambda \hat B_k]$, and  $\langle A_j\rangle_\lambda$ is an arbitrary function 
of $\lambda$, bounded by the maximum and minimum of the set of the allowed values of $A_j$. If no such LHS model exists then Alice is said to be able to steer Bob's system via her measurements. 

We restrict our attention to the case where all outcomes are labelled by $\pm 1$, and Bob's measurements correspond to a set of orthogonal spin directions on a qubit Hilbert space, i.e., $\hat B_k ={\bs b}_{k}\cdot \hat {\bs{\sigma}}$ with ${\bs b}_{k}\cdot {\bs b}_{k'}=\delta_{kk'}$. 
Here $\hat{\bs \sigma}=(\hat\sigma_1,\hat\sigma_2,\hat\sigma_3)$ is the vector of Pauli operators in some fixed basis. 
It is shown in the Supplemental Material~\cite{SM} that any LHS model for this case must satisfy the steering inequality
\begin{equation} \label{steer1}
\|M\|_{\rm tr}:= {\rm tr}\sqrt{M^\top  M} \leq \sqrt{m}. 
\end{equation}
That is, there is an experimentally measurable \textit{steering parameter} (here, the trace-norm of the correlation matrix $M$) which, for a LHS model, 
has an upper \textit{bound} independent of the results 
(here, the square root of the number of Alice's settings). Thus, violation of this inequality is sufficient for Alice to be able to steer Bob.

Suppose now that Alice and Bob genuinely wish to achieve violation of steering inequality (\ref{steer1}) for some shared two-qubit state $\hat\rho$, by each choosing a set of mutually orthogonal measurement directions. Thus, Alice measures a set of spin operators $\hat A_j:={\bs a}_{j}\cdot{\bs \sigma}$ with ${\bs a}_{j}\cdot {\bs a}_{j'}=\delta_{jj'}$, and  $M_{jk}={\bs a}_{j}^{\top}T{\bs b}_{k}$, where $T$ is the $3\times3$ spin correlation matrix for state $\hat\rho$, i.e., $T_{pq}:={\rm Tr}[\hat\rho\hat\sigma_p\otimes\hat\sigma_q]$. The steering parameter in Eq.~(\ref{steer1}) is then predicted to be~\cite{SM}
\begin{equation} \label{steer2}
\|M\|_{\rm tr} = \blk \|P_ATP_B\|_{\rm tr} %\leq \sqrt{m}},
\end{equation}
where $P_A:=\sum_j{\bs a}_{j}{\bs a}_{j}^{\top}$ 
    and $P_B:=\sum_k{\bs b}_{k}{\bs b}_{k}^{\top}$ are the respective $3\times 3$ projection matrices onto the span of Alice's and Bobs measurement directions.

If Alice and Bob each choose a {\it triad} of mutually orthogonal directions, i.e., $m=n=3$, then $P_A=P_B=I_3$ and Eq.~(\ref{steer2}) simplifies to ${\rm  tr}\sqrt{T^\top  T}$, independently of the particular triads chosen.  Thus, the degree of steerability, as quantified by a violation of \erf{steer1}, is invariant under local rotations, and so can be established even when Alice and Bob do not share any reference directions. In particular, for a Werner state \textemdash a probabilistic mixture of a maximally entangled singlet state with a symmetric noise state parametrised by the mixing probability, or Werner parameter, $W$\textemdash one has $T=-WI_3$, implying that a (constant) violation is guaranteed for any $W>1/\sqrt{3}$.  In comparison, a corresponding violation of the Bell inequalities in Refs.~\cite{Wallman2011,Shadboldt2012,Palsson2012} is only guaranteed for $W=1$, and the degree of violation can be arbitrarily small.

For the case where Alice and Bob each choose a {\it pair} of mutually orthogonal directions, i.e., $m=n=2$, $P_A$ and $P_B$ are the projections onto the planes spanned by their measurement directions. Hence, the corresponding degree of steerability witnessed by the steering inequality is invariant under any local rotations that leave the measurement directions within these planes. In particular, if Alice and Bob only share a single reference direction $r$, then they can determine a degree of steerability invariant under arbitrary rotations about this direction, by choosing their measurement directions to lie in the plane orthogonal to $r$.  For a Werner state, violation is guaranteed for any $W>1/\sqrt{2}$ (the best possible bound for this case \cite{Cavalcanti2015}). In comparison, a violation of the Bell inequalities in Refs.~\cite{Wallman2011,Shadboldt2012,Palsson2012} is again only guaranteed for $W=1$, and may be arbitrarily small.

{\it Necessary and sufficient steering (NSS) inequality.--} For $m=n=2$, it is of interest to compare the RIS inequality~(\ref{steer1}) with a recent necessary and, for the case of maximally mixed marginals, sufficient condition for the correlation matrix $M$ to admit a qubit LHS model for Bob~\cite{Cavalcanti2015}:
\begin{equation} \label{josab1}
	|M^\top u_+| + |M^\top u_-| \leq \sqrt{2}, 
\end{equation} 
with $u_\pm :=(1,\pm1)^\top /\sqrt{2} $. Note that we have normalised \erf{josab1} differently from the inequality in Ref.~\cite{Cavalcanti2015} so that it has the same bound as \erf{steer1} for $m=2$. If Alice and Bob share a two-qubit state $\hat\rho$, and Alice measures in two orthogonal directions ${\bs a}^{(1)}$ and ${\bs a}^{(2)}$, 
the predicted steering parameter in \erf{josab1} reduces to~\cite{SM} 
\begin{equation} \label{josab2}
|M^\top u_+| + |M^\top u_-|=|P_BT^\top \bs a_+| + |P_BT^\top \bs a_-|  ,  %\leq 2.
\end{equation}	
 with $\bs a_\pm=({\bs a}^{(1)}\pm {\bs a}^{(2)})/\sqrt{2}$. Thus, unlike the RIS inequality, the NSS inequality is not invariant under rotations in the plane of Alice's measurement directions. However, minimising Eq.~(\ref{josab2}) over all such rotations  recovers  Eq.~(\ref{steer2}) (for $m=2$)~\cite{SM}.
In this sense, our RIS inequality (\ref{steer1}) is the best possible for $m=n=2$, and we conjecture that it is similarly optimal for $m=n=3$.

As an example of practical interest, let $\Phi$ denote the angle between Alice and Bob's measurement planes, and $\alpha$ denote the angle that the line of intersection of these planes makes with Alice's measurement direction ${\bs a}^{(1)}$.  For a Werner state, with $T=-WI_3$, the RIS parameter of~\erf{steer2} becomes
\begin{equation}\label{steer3}
\|M\|_{\rm tr}= W \left(1+|\cos\Phi| \right), 
\end{equation}
independently of $\alpha$, while the NSS parameter of \erf{josab1}, becomes 
\begin{align} 
|M^\top u_+| +& |M^\top u_-|=
 W \left(\sqrt{1+\cos^{2}\Phi+\sin 2\alpha \sin^{2}\Phi}\right. \nonumber \\
&+\left.\sqrt{1+\cos^{2}\Phi-\sin 2\alpha \sin^{2}\Phi}\, \right)/\sqrt{2}.  \label{josab3}
\end{align}
Minimising this over $\alpha$ recovers Eq.~(\ref{steer3}).

\begin{figure}[htbp]
\includegraphics[width=0.48\textwidth]{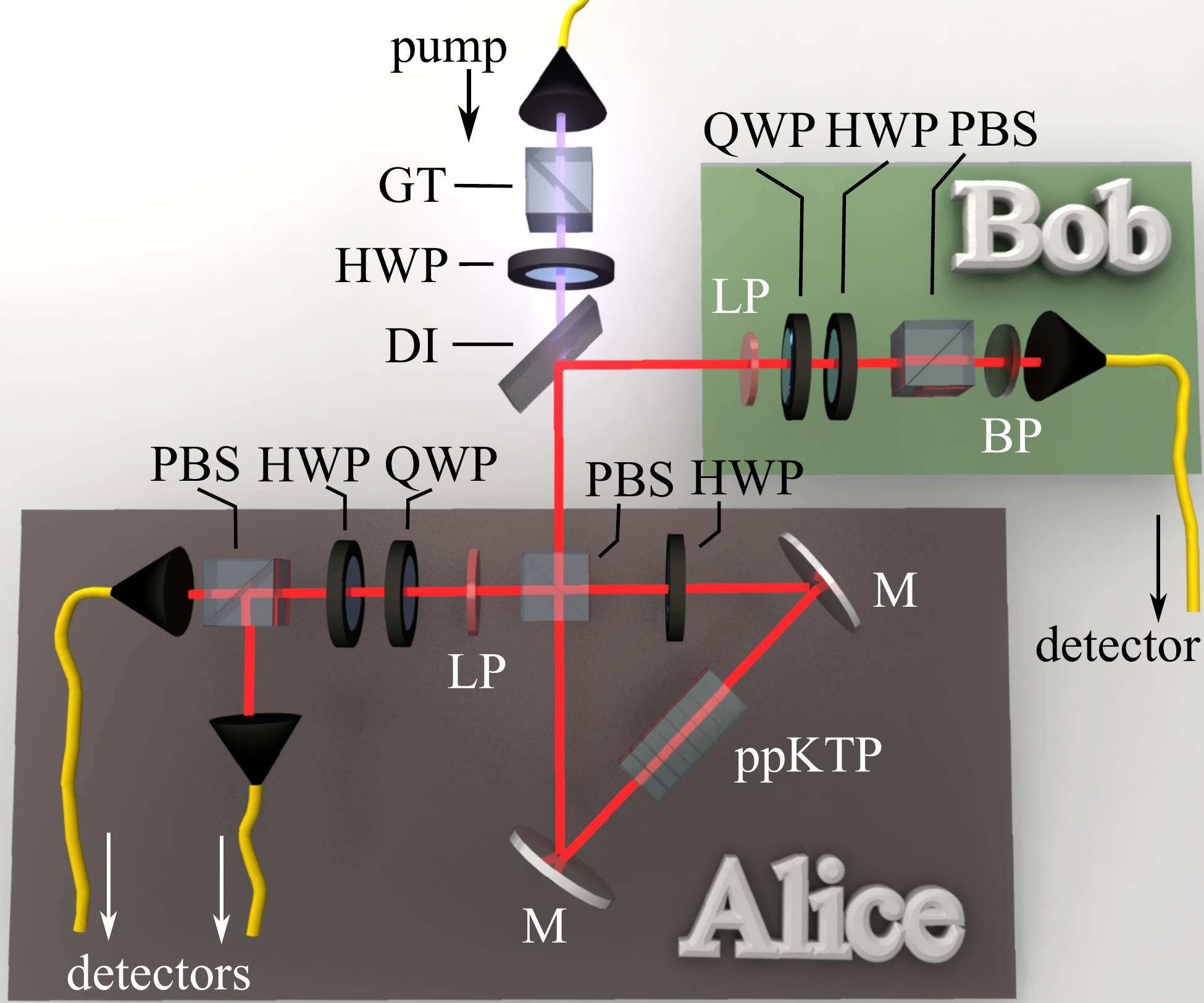}
\caption{
Entangled photon pairs at 820 nm were produced via SPDC in a Sagnac interferometer. Different measurement settings are performed by rotating half- and quarter-wave plates (HWP and QWP) relative to the polarizing beam splitters (PBS). Long pass (LP) filters and an additional bandpass filter in Bob's line, remove 410 nm pump photons co-propagating with the 820 nm photons, before photons are coupled into single-mode fibres and detected by single photon counting modules and counting electronics.}
\label{setup}
\end{figure}

{\it Experimental setup.--}
As shown in Fig.~\ref{setup}, we implemented these steering protocols using polarization-entangled states generated from a spontaneous parametric down-conversion (SPDC) source. A 10 mm-long periodically poled potassium titanyl phosphate (ppKTP) crystal, mounted in a polarization Sagnac ring interferometer~\cite{Kim2006,Fedrizzi2007}, was pumped bidirectionally by a 410 nm fibre-coupled continuous-wave laser with an output power (after fibre) of 2.5 mW.

To test the quality of the generated entangled state, quantum state tomography~\cite{White2007} was performed at several stages throughout the experiment---in each case, we achieved a fidelity of {\em ca.}~$98\%$ with the singlet state $\left(\ket{HV}-\ket{VH}\right)/\sqrt{2}$. We measured the correlations in our experiment by rotating the QWPs and HWPs in front of polarising elements to set measurement directions and implement projective measurements for Alice’s $m$ and Bob’s $n$ settings, and counted coincident detections. We calculated each steering parameter from the observed correlations, and determined its error from those in the correlation matrix elements: 
$\Delta M_{jk}=\sqrt{(\Delta M_{jk}^{\textrm{(sys)}})^2+(\Delta M_{jk}^{\textrm{(stat)}})^2}$. The error consists of a systematic error due to small imperfections in Bob's measurement settings, which could lead to an overestimation of the correlations~\cite{Bennet2012}, and the statistical error arising from Poissonian statistics in photon counting.
Quantum steering usually requires that Bob chooses his settings independently from one measurement to the other. However, as we control Alice’s realization and apparatus in this demonstration (i.e.\ we are not in a adversarial scenario), there is no need for a time ordering of the events and we collected data without shot-to-shot randomization~\citep{Bennet2012}. However, this would have to be altered in a full deployment~\cite{Wittmann2012}.

\begin{figure}[htbp]
\includegraphics[width=0.47\textwidth]{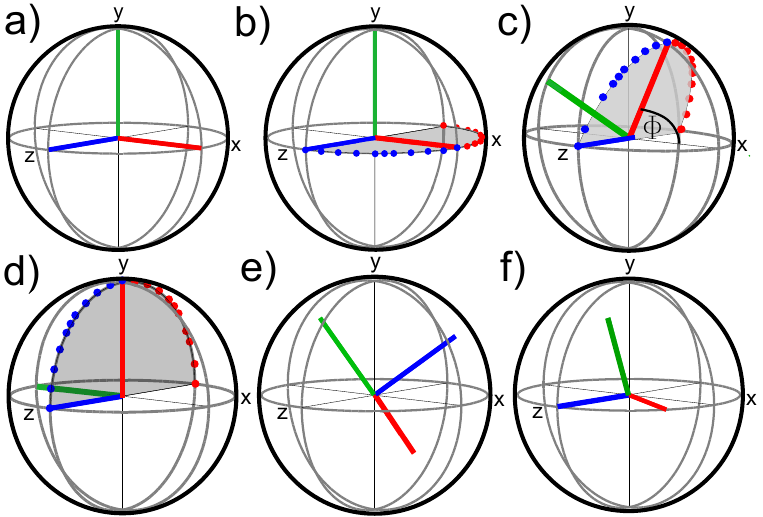}
\caption{The Poincar\'{e} (Bloch) spheres contain vectors showing one of the eigenstates of the three relevant directions (blue, red and green) in the experiments we performed. In each of the experiments, measurements are made along two or three of these directions. (a) Bob uses the same three measurement directions in the each of the $n=3$ experiments, while using only the red and blue directions for experiments with $n=2$. (b) Alice's directions, in the case where Alice and Bob share a reference direction (green). We test the invariance of $m=2$ measurements to rotations in the plane (grey), as the blue and red settings are rotated through $90{^\circ}$ in steps (blue and red dots). $\Phi=0^{\circ}$ denotes the fact that the plane is not tilted with respect to Bob, c.f.\ next panels. (c) Alice's directions for $m=2$ (blue and red dots) when her plane of measurement directions is tilted by $\Phi=64^{\circ}$ and the settings are rotated in that plane, whilst maintaining local orthogonality. (d) Same as (c), but with $\Phi=90^{\circ}$. (e) Alice's orthogonal measurement directions for the $m=n=3$ experiment are strongly misaligned with respect to Bob \cite{SM}. (f) Nonorthogonal measurement directions for Alice \cite{SM}.}
\label{Bloch}
\end{figure}

\textit{Experimental tests and results.---}
We investigated the rotational invariance of quantum steering in a series of experiments.\\

\noindent Case 1. We first considered the case where Alice and Bob share a single reference direction and use $m=n=2$ measurement settings---the minimal set size. The measurement directions lie in a plane orthogonal (on the Bloch sphere) to the shared direction, and the two settings on each side are locally orthogonal. This is a natural physical situation because a shared reference direction may be determined reliably, for example, by line of sight between the parties. Furthermore, it is natural to assume that Alice and Bob can reliably set local measurement directions. However, although Alice and Bob's measurement directions will lie in the same plane, their relative orientation within this plane may be unknown. This situation also provides for a direct comparison between the RIS and NSS inequalities.

\begin{figure}[htbp]
\includegraphics[width=0.45\textwidth]{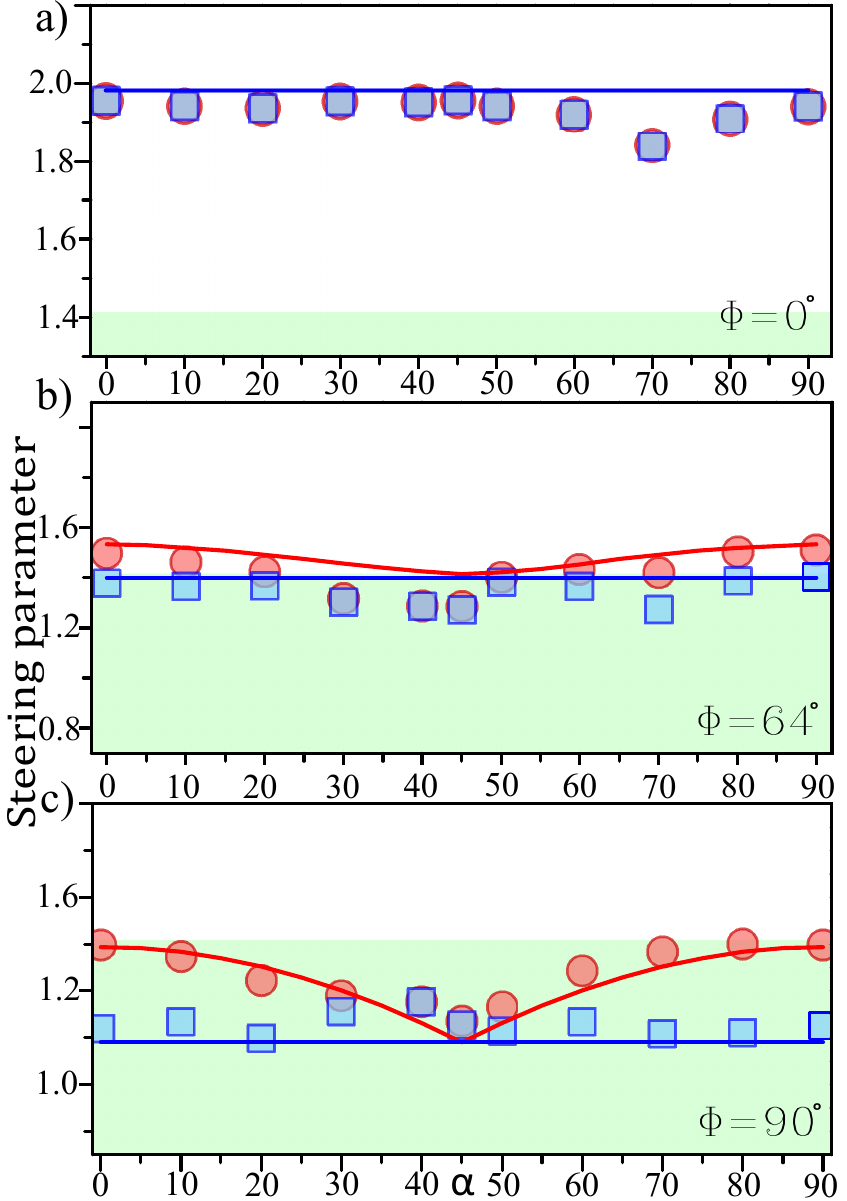}
\caption{The steering parameter versus rotation angle $\alpha$ in Alice's measurement plane, for experiments with $m=n=2$ measurement directions (Cases 1 and 2 of the main text).
The plane tilt angle takes on the values (a) $\Phi=0^{\circ}$, (b) $\Phi=64^{\circ}$, (c) $\Phi=90^{\circ}$ (see Fig.~\ref{Bloch}(b)-(d) respectively). For all angles, we calculated the theoretically expected curves for the RIS inequality~(\ref{steer1}) (blue) and the NSS inequality~(\ref{josab1}) (red) for the Werner state (solid line). The Werner parameter $W$ of the closest Werner state was calculated from the average tomographic data.
The RIS data is represented by blue squares, and the NSS data by red circles. The error bars are too small to be seen. Data points in the upper white region imply steering of Bob by Alice.}
\label{rotation}
\end{figure}

In our experiment, the measurements lie in the $\sigma_{x}$-$\sigma_{z}$ plane (Fig.~\ref{Bloch} a,b), corresponding to an angle of $\Phi=0^{\circ}$ between Alice and Bob's measurement planes. While Bob's measurement directions were kept constant, Alice's were rotated through 90$^{\circ}$ in the plane, by angles $\alpha\in\{0^{\circ},10^{\circ},20^{\circ},30^{\circ},40^{\circ},45^{\circ},50^{\circ},60^{\circ},70^{\circ},80^{\circ},90^{\circ}\}$.
We observed a rotation-independent violation of both the RIS and NIS inequalities for $\Phi=0^{\circ}$ (Figs.~\ref{Bloch}b and \ref{rotation}a), except for some deviation around $\alpha=70^{\circ}$. Across the remainder of the range, the measured steering parameters are close to the theoretically-predicted value of $1.97$ (Fig.\ref{rotation}a solid line), for both the RIS and NSS correlation functions of a Werner state~\cite{Werner2001} having the same fidelity with the singlet as our entangled state. This value is close to the maximum value of $2$ for an ideal singlet state. We attribute the experimental deviation near $\alpha=70^{\circ}$ to time-dependent fluctuations of the end state due to temperature shifts affecting the source.
 While this imperfection is undesirable, it serves to illustrate the point that the RIS inequality is tolerant to noise, due to the large gap at all relative angles $\alpha$ between the bound of $\sqrt{2}$ in Eq.~(\ref{steer1}) and the theoretical maximum value of $2$.
 
\noindent Case 2. We repeated case 1, but allowed an offset in the previously-shared reference direction (i.e.\ $\Phi \neq 0$), simulating the case when there is imperfection in sharing this direction.
Due to robustness of the inequalities, we had to tilt Alice's measurement plane significantly, by $64^{\circ}$, to shift to a regime where the inequalities were not necessarily violated (Fig.~\ref{Bloch}c).
The RIS data stayed approximately rotation-invariant (Fig.~\ref{rotation}b), and comparable to the theoretically-predicted value of $1.40$ in Eq.~(\ref{steer3}) (less than the steering bound of $\sqrt{2}$) 
for the closest ideal Werner state. 
By contrast, the NSS data showed an oscillatory behaviour (Fig.~\ref{rotation}b dashed line), as predicted by Eq.~(\ref{josab3}), with violation for $\alpha<20^{\circ}$ and $\alpha>70^{\circ}$. Again, the noise in the data is due to asymmetries in the state arising from state preparation imperfections due to thermal fluctuations in the apparatus. 

We also investigated the case where there was extreme misalignment in the supposedly-shared reference direction. For this, Alice used measurement directions in the  $\sigma_{z}$-$\sigma_{y}$ plane, i.e.\ for $\Phi=90^{\circ}$ (Fig.~\ref{Bloch}d). Neither steering inequality was violated at any angle $\alpha$.
While the RIS data was approximately insensitive to rotations of Alice's measurement directions, the NIS data showed an oscillatory behaviour (Fig.\ref{rotation}c), as per the theoretical predictions in Eqs.~(\ref{steer3}) and~(\ref{josab3}) respectively for the tomographically-reconstructed state. For each of Fig.~(\ref{rotation}a)-(\ref{rotation}c) the RIS values are never greater than the NSS values, as predicted.

\noindent Case 3. As the RIS inequality~(\ref{steer1}) is not restricted to $m=n=2$, we extended the number of measurement directions to $m=n=3$ directions for each party. First we studied the case where Alice and Bob's orthogonal measurement triads were perfectly aligned, along the $\sigma_x$, $\sigma_y$ and $\sigma_z$ directions (Fig.~\ref{Bloch}a). For this case, we generated a state with a fidelity of $98.4\%$ with a singlet state. The measured RIS steering parameter $2.93\pm0.01$ significantly exceeds the bound of $\sqrt{3}$ in inequality (\ref{steer1}). The  small deviation from the maximum possible value of 2.95 for a Werner state with W=0.984 can be explained by imperfections of polarisation optics and classical interference in the Sagnac interferometer. 
We also calculated the average correlation between Alice's and Bob's result for $m\neq n$ directions. We analysed a subset of the 3-setting-per-side data to investigate if Alice is able to steer Bob's state for $m=2$ ($\sigma_x$ and $\sigma_z$) and $n=3$, and for $m=3$ and $n=2$ measurement directions. In both cases the RIS inequality bound of $\sqrt{3}$ was violated, with respective steering parameters $1.96\pm0.01$ and $1.97\pm0.01$. 
Finally, we investigated the case of miscalibration of Alice's measurement directions. We chose an orthogonal triad for Alice (Fig.~\ref{Bloch}e) which was strongly misaligned with Bob's measurement directions (Fig.~\ref{Bloch}a)~\cite{SM}. 
The measured steering parameter was $2.21\pm0.01$, well above the bound of $\sqrt{3}$ in Eq.~(\ref{steer1}). This particular measurement was conducted with a state having slightly lower fidelity ($96\%$) with a singlet state, due to environmental fluctuations in the laboratory. We note that this orientation of measurement directions would not lead to a steering demonstration using an ordinary linear steering inequality of the type given in Ref.~\citep{Saunders2010}, but the RIS is robust to such major misalignments.\\

\noindent Case 4. Finally, we observed whether Alice could demonstrate steering by measuring in nonorthogonal directions, while Bob's measured directions remained orthogonal (Fig.\ref{Bloch}f). We note that that the steering inequalities (Eq.~2 and~4) are valid for any choice of Alice's measurement directions, whereas the predictions for the steering paramters in Eq.~3-7 assume they are orthogonal. First, we considered the case of two measurement directions for each party. Bob measured along $\sigma_{x}$ and $\sigma_{z}$ (Fig.~\ref{Bloch}a red and blue), with Alice's directions along  $\sigma_{z}$ and at a $60^{\circ}$ angle  therefrom in the same plane (Fig.~\ref{Bloch}f blue and red). We generated a state with a fidelity of $97.2\%$. With these measurement settings, both the RIS and NSS inequalities, \erfs{steer1}{josab1},
were violated, with steering parameters $1.85\pm0.01$ and $1.96\pm0.01$ 
compared to the bound of $\sqrt{2}$.
We concluded the experiment by measuring in three directions for each party. While Bob measured along $\sigma_{x}$, $\sigma_{y}$ and $\sigma_{z}$(Fig.~\ref{Bloch}a), Alice's directions formed a regular tetrahedron with the origin (Fig.~\ref{Bloch}f) \cite{SM}. We violated the bound of $\sqrt{3}$ in Eq.~(\ref{steer1}) with a steering parameter of $2.74\pm0.01$, which is comparable to the maximum possible value of 3 obtainable via mutually orthogonal directions and a maximally entangled state.

{\it Conclusions.---} 
We theoretically determined a rotation-invariant steering inequality. Sufficiently entangled Werner states produce constant violations of the inequality under local rotations. For two measurement settings per side, we showed that the violation is constant under local rotations about a shared axis and that our RIS inequality is the optimal such inequality for this situation. Experimentally, we showed that, for two settings per side and one shared reference direction only, the violation of both inequalities are independent of frame alignment between Alice and Bob, up to state preparation imperfections. Degradation of the shared direction eventually means that steering inequalities can no longer be violated. For three settings per side, the rotationally-invariant inequality is violated even for maximal misalignment of the reference frames, unlike an ordinary steering inequality~\cite{Saunders2010} and even in the presence of state preparation imperfections. In principle, using the appropriate (two- or three-setting) rotation-invariant inequality for one or zero shared measurement directions always provides a large buffer between the theoretically-expected steering value and the bound, unlike the case for frame rotations in Bell tests~\cite{Shadboldt2012,Palsson2012}. As demonstrated by our data, this provides robustness to imperfections such as asymmetries in a real-world shared entangled state. Therefore our work shows how the steering task can be more tolerant to reference-frame misalignment and asymmetry than Bell tests, adding to the previous list (decoherence-tolerance~\cite{Saunders2010} and loss-tolerance~\cite{Bennet2012}) of noise sources where steering enjoys an advantage. Our demonstration of rotationally-invariant steering holds potential application in ground-to-space satellite quantum communication~\cite{Vallone2015} and in quantum key distribution~\cite{Slater2014}. \\

We acknowledge Sergei Slussarenko for helpful discussions, and thank Cyril Branciard for useful comments, including on the role of maximally mixed marginals in the NSS condition.
This work was supported by ARC project DP140100648.

\bibliographystyle{apsrev}

\begin{thebibliography}{22}
\bibitem[{sup()}]{referred}
\bibinfo{note}{The exception is quantum-refereed games \cite{Kocsis2015,Cavalcanti2013,Buscemi2012,verbanis,Branciard2013} where the need for trust can be mitigated by using additional quantum resources}.
\end{thebibliography}

\newpage

\appendix{SUPPLEMENTAL MATERIAL}

\section{Mathematical details}\label{supp_math}

Here we provide the necessary details underlying the steering inequalities in the main text.

To prove the rotationally-invariant steering inequality in Eq.~(\ref{steer1}), suppose that the measurement outcomes for Alice and Bob are restricted to $\pm 1$, and that the correlation matrix $M$ has a qubit LHS model for Bob's system as per Eq.~(\ref{lhs}) of the main text.  Thus,
\begin{equation} \label{lhssupp}
	M_{jk}= \langle A_jB_k\rangle =\int d\lambda \,p(\lambda) \,\langle A_j\rangle_\lambda\,\langle \hat B_k\rangle_{\hat \varrho_\lambda },
\end{equation} 
with  $\langle \hat B_k\rangle_{\hat \varrho_\lambda }= {\rm Tr}[\hat \varrho_\lambda \hat B_k]$ where $$\hat \varrho_\lambda =\frac{1}{2}[1+\bs{s}(\lambda)\cdot\hat{\bs{\sigma}}],~~\hat B_k ={\bs b}_{k}\cdot \hat{\bs{\sigma}},~~{\bs b}_{k}\cdot {\bs b}_{k'}\blk =\delta_{kk'}.$$
Hence, letting ${\cal A}(\lambda)$ denote the $m$-vector with components ${\cal A}_j(\lambda)=\langle A_j\rangle_\lambda$, and ${\cal B}(\lambda)$ denote the $n$-vector with components ${\cal B}_{k}(\lambda)={\bs s}(\lambda)\cdot {\bs b}_{k}$, one can rewrite the correlation matrix in Eq.~(\ref{lhssupp}) as
\begin{equation}
M = \int d\lambda \,p(\lambda)\,{\cal A}(\lambda)\,{\cal B}(\lambda)^\top .
\end{equation}
Taking the trace norm then yields the steering inequality
\begin{align}
\|M\|_{\rm  tr} &= \left\|\int d\lambda \,p(\lambda)\,{\cal A}(\lambda) \,{\cal B}(\lambda)^\top  \right\|_{\rm tr} \nn \\
&\leq \int d\lambda \,p(\lambda)\,\| {\cal A}(\lambda) \,{\cal B}(\lambda)^\top  \|_{\rm tr}\nn\\
&= \int d\lambda \,p(\lambda)\,| {\cal A}(\lambda)|\,|{\cal B}(\lambda)|\nn\\
&\leq  \int d\lambda \,p(\lambda)\,\sqrt{m} = \sqrt{m} ,\label{gen}
\end{align}
as per Eq.~(\ref{steer1}) of the main text.
Here, the first inequality follows from the triangle inequality, the next line from the easily verified property  $\|vw^\top \|_{\rm tr}=|v|\,|w|$, and the final inequality via $|{\cal A}(\lambda)|^2 = \sum_j \langle A_j\rangle_\lambda^2\leq m$ and $|{\cal B}(\lambda)|^2 =\sum_k {\bs s}(\lambda)^\top  {\bs b}_{k}{\bs b}_{k}^{\top}{\bs s}(\lambda) = {\bs s}(\lambda)^\top  P_B{\bs s}(\lambda) =|P_B{\bs s}(\lambda)|\grn ^2 \blk\leq |{\bs s}(\lambda)|\leq 1$.

Now, if Alice and Bob each make a set of mutually orthogonal measurements on a two-qubit state with spin correlation matrix $T$, with $T_{jk}={\rm Tr}[\hat\rho\, \sigma_j\otimes\sigma_k]$, then 
\begin{equation}
	M = A^\top TB,
\end{equation}
where $A$ and $B$ denote the $3\times m$ and $3\times n$ matrices with columns corresponding to their respective spin directions, i.e., $A:=({\bs a}_1~{\bs a}_2\dots {\bs a}_m)$ and $B:=({\bs b}_1\dots {\bs b}_n)$.  The trace norm of $M$ can then be rewritten as
\begin{align}
\|M\|_{\rm tr} &=\|M^\top \|_{\rm tr} = {\rm Tr}\sqrt{A^\top TBB^\top T^\top A}\nn\\
&={\rm Tr}\sqrt{A^\top TP_BP_BT^\top A} \nn\\
&=\|P_BT^\top A\|_{\rm tr} =\|(P_BT^\top A)^\top \|_{\rm tr} \nn\\
&= {\rm Tr}\sqrt{P_BT^\top AA^\top  TP_B} \nn\\
&= {\rm Tr}\sqrt{P_BT^\top P_AP_A TP_B} = \|P_ATP_B\|_{\rm tr},
\end{align}
 as per Eq.~(\ref{steer2}) of the main text, where we have used $BB^\top =\sum_j {\bs b}_{j} {\bs b}_{j}^{\top} = P_B=P_B^2$, and the corresponding relations for $AA^\top $. 

To obtain Eq.~(\ref{josab2}) of the main text, from the NSS inequality 
\begin{equation} \label{josabsupp1}
	|M^\top u_+| + |M^\top u_-| \leq \sqrt{2}
\end{equation} 
in Eq.~(\ref{josab1}) of the main text, note for any 2-vector $u$ that $|M^\top u|^2=u^\top MM^\top u= u^\top A^\top TBB^\top T^\top Au= u^\top A^\top TP_BP_BT^\top Au = |P_BT^\top Au|^2$.  Substitution into the above inequality, and recalling that $u_\pm=(1,\pm1)^\top /\sqrt{2}$ and $\bs a_\pm=({\bs a}_1\pm{\bs a}_2)/\sqrt{2}$, immediately yields
\begin{equation} \label{josab2supp}
|M^\top u_+| + |M^\top u_-|=|P_BT^\top {\bs a}_+| + |P_BT^\top {\bs a}_-|
\end{equation}
 for the NSS steering parameter, as required.  

Finally, we show that, for $m=n=2$, the RIS steering parameter in Eq.~(\ref{steer2}) is given by minimising the NIS steering parameter in Eq.~(\ref{josab2}), over all orthogonal measurement pairs in Alice and Bob's respective measurement planes. Noting that Eq.~(\ref{josab2}) (equivalent to Eq.~(\ref{josab2supp}) above) only depends on Bob's measurement directions via $P_B$, it is sufficient to show that 
\begin{equation} \label{suffsupp}
\min_{R_A} |P_BT^\top R_A{\bs a}_+| + |P_BT^\top R_A{\bs a}_-| = \|P_ATP_B\|_{\rm tr},
\end{equation}
where $R_A$ ranges over all rotations that leave Alice's measurement plane invariant and $a_+$ and $a_-$ are fixed.  
Now, for any vector ${\bs a}$ in this measurement plane one has $P_A{\bs a}={\bs a}$, and hence, using $P_B^2=P_B$, 
\begin{equation}
|P_BT^\top {\bs a}|=|P_BT^\top P_A{\bs a}| = \sqrt{\bs a^\top P_ATP_BT^\top P_A{\bs a}} .
\end{equation}
Defining $K:=P_ATP_BT^\top P_A$, one therefore has
\begin{align*}
|P_BT^\top &R_A{\bs a}_+| + |P_BT^\top R_A{\bs a}_-|\\ &= \sqrt{(R_A{\bs a}_+)^\top K(R{\bs a}_+)}
+ \sqrt{(R_A{\bs a}_-)^\top K(R{\bs a}_-)}.
\end{align*}
Since $R_A{\bs a}_\pm$ and $K$ only have support on Alice's measurement plane, minimising this expression over $R_A$ reduces to a $2\times2$ matrix problem.	Further, since $K$ is by definition a nonnegative symmetric matrix, and $R_A{\bs a}_+$ and $R_A{\bs a}_-$ range over all pairs of orthogonal  unit  vectors in the measurement plane, we can choose coordinates such that 
\begin{equation}
K\equiv \left( \begin{array}{cc} k&0\\0& k'\end{array}  \right),~R_A{\bs a}_+\equiv\left( \begin{array}{c} \cos\theta\\ \sin\theta\end{array}  \right),~R_A{\bs a}_-\equiv\left( \begin{array}{c} \sin\theta\\ -\cos\theta\end{array}  \right) \nn
\end{equation}
on this plane, for some $k\geq k'\geq0$ and $\theta\in[0,2\pi]$. Thus,
\begin{align}
|P_BT^\top &R_A{\bs a}_+| + |P_BT^\top R_BR_A{\bs a}_-| \nn\\
&= \sqrt{k\cos^2\theta+k'\sin^2\theta} 
+\sqrt{k\sin^2\theta+k'\cos^2\theta} \nn\\
&=\sqrt{X+Y\cos2\theta} +\sqrt{X-Y\cos2\theta},
\end{align}
with $X:=(k+k')/2$ and $Y:=(k-k')/2$.  It is straightforward to check that the function $f(x):=\sqrt{1+x}+\sqrt{1-x}$ is symmetric with a single maximum at $x=0$.  Hence, the minimum of the above expression is obtained at $\cos 2\theta=\pm1$, yielding
\begin{align}
\min_{R_A} 	|P_BT^\top &R_A{\bs a}_+| + |P_BT^\top R_BR_A{\bs a}_-| \nn\\
&=\sqrt{X+Y}+\sqrt{X-Y} = \sqrt{k}+\sqrt{k'}\nn\\
&= {\rm Tr}[\sqrt{K}]=\| P_BT^\top P_A\|_{\rm  tr} ,
\end{align}
using the definition of $K$. Finally, since the trace norm of a matrix is invariant under transposition, Eq.~(\ref{suffsupp}) follows as desired.

Substantial generalisations of these results, with Alice and Bob not limited to orthogonal sets of measurements, and allowing for detector inefficiencies, will be discussed elsewhere.

\section{Measurement directions}\label{supp_directions}
To demonstrate quantum steering for the $m=n=3$ for strongly misaligned measurement directions (Fig.~2e) in case 3 of the main text, we chose Bob to measure along $\sigma_{x}$, $\sigma_{y}$ and $\sigma_{z}$. Alice's directions are:
\begin{equation}
\begin{aligned}
 \bs a_{1}&=(\frac{1}{\sqrt{3}},\frac{1}{\sqrt{3}},\frac{1}{\sqrt{3}}), \\
 \bs a_2&=(\frac{(1+\sqrt{3})}{\sqrt{12}},\frac{-2}{\sqrt{12}},\frac{(1-\sqrt{3})}{\sqrt{12}}),\\
 \bs a_3&=(\frac{(1-\sqrt{3})}{\sqrt{12}},\frac{-2}{\sqrt{12}},\frac{(1+\sqrt{3})}{\sqrt{12}}).
 \end{aligned}
 \end{equation}
For showing quantum steering for nonorthogonal $m=n=3$ measurement directions in case 4 of the main text (Fig.~2f), we chose Alice's directions as,
\begin{equation}
\begin{aligned}
 \bs a_{1}&=(1,0,0), \\
 \bs a_2&=(\frac{1}{2},\frac{1}{2\sqrt{3}},\sqrt{\frac{2}{3}}),\\
 \bs a_3&=(\frac{1}{2},\frac{\sqrt{3}}{2},0).
 \end{aligned}
 \end{equation}
We chose Bob to measure along $\sigma_{x}$, $\sigma_{y}$ and $\sigma_{z}$.

\end{document}